\documentclass[twocolumn,prb,showpacs,byrevtex,superbib]{revtex4}
\usepackage{amsmath}
\usepackage{graphicx}
\usepackage{epsfig}
\usepackage[bf]{subfigure}
\usepackage[scaled=.92]{helvet}
\usepackage{courier}
\usepackage{mathptmx}

\input{tcilatex}
\DeclareMathAlphabet{\mathsf}{OT1}{cmss}{bx}{n}

\begin{document}

\title
[Travel time Stability]{\Large{Travel time stability in weakly range-dependent sound channels}}%

\author{\textsf{F. J. Beron-Vera}$^{\textrm{a})}$\footnotetext{$^{\textrm{a})}$Author to whom correspondence should be addressed. Electronic mail: \texttt{fberon@rsmas.miami.edu}}}%

\affiliation{RSMAS/AMP, University of Miami, Miami, FL 33149}%

\author{\textsf{M. G. Brown}}%

\affiliation{RSMAS/AMP, University of Miami, Miami, FL 33149}%



\begin{abstract}%

Travel time stability is investigated in environments consisting of a
range-independent background sound-speed profile on which a highly
structured range-dependent perturbation is superimposed. The stability of
both unconstrained and constrained (eigenray) travel times are considered.
Both general theoretical arguments and analytical estimates of time spreads
suggest that travel time stability is largely controlled by a property $%
\omega ^{\prime }$ of the background sound speed profile. Here, $2\pi
/\omega (I)$ is the range of a ray double loop and $I$ is the ray action
variable. Numerical results for both volume scattering by internal waves in
deep ocean environments and rough surface scattering in upward refracting
environments are shown to confirm the expectation that travel time stability
is largely controlled by $\omega ^{\prime }$.\newline

\end{abstract}%

\volumeyear{2002}%

\volumenumber{205}%

\issuenumber{5}%

\date[Dated: ]{\today}%

\startpage{1}%

\endpage{102}%

\eid{identifier}%

\maketitle%

\section{\textsf{Introduction}}

Measurements made during the Slice89 propagation experiment \cite%
{Duda-etal-92}, performed in the eastern North Pacific Ocean, suggested that
in that environment the near-axial energy is more strongly scattered than
the energy corresponding to the steeper rays. Similar behavior was
subsequently observed in measurements made during the AET experiment \cite%
{Worcester-etal-99,Colosi-etal-99}, also performed in the eastern North
Pacific Ocean. Motivated in large part by these observations, several
studies \cite%
{Duda-Bowlin-94,Simmen-Flatte-Yu-Wang-97,Smirnov-Virovlyansky-Zaslavsky-01,Beron-Brown-03}
have been carried out to investigate the dependence of ray stability on
environmental parameters. All of these studies have focused on ray stability
in either physical space (depth, range) or phase space (depth, angle). The
present study extends the earlier work by considering the sensitivity of ray
travel times to environmental parameters. Our focus on travel times is more
closely linked to the Slice89 and AET observations than the earlier
sensitivity studies inasmuch as both sets of measurements were made in depth
and time at a fixed range.

In this study we consider ray motion in environments consisting of a
range-independent background on which a range-dependent perturbation, such
as that produced by internal waves in deep ocean environments, is
superimposed. We consider the influence of the background sound speed
structure on both unconstrained and constrained (eigenray) measures of
travel time spreads. Surprisingly, the conclusion of this work is that
travel time stability is largely controlled by the same property \cite%
{Beron-Brown-03} of the background sound speed structure that controls ray
trajectory stability. Although this work was motivated by the deep ocean
measurements mentioned above, the results presented apply to a much larger
class of problems. To illustrate this generality we include numerical
simulations of ray scattering by a rough surface in upward refracting
environments.

The remainder of the paper is organized as follows. In Sec. \ref{TheoBack}
the equations on which our analysis is based are presented. These are the
coupled ray/travel time equations written in terms of both the usual phase
space variables and action--angle variables. In Sec. \ref{UTimeSpread} two
simple expressions for unconstrained travel time spreads in terms of
action--angle variables are derived (trivially) and shown to be in good
agreement with numerical simulations. In Sec. \ref{CTimeSpread} an
expression for constrained (eigenray) time spreads is derived, again using
action--angle variables, and shown to be in good agreement with numerical
simulations. The same result was recently derived using a different argument
by Virovlyansky \cite{Virovlansky-03}. The combination of the results
presented in Secs. \ref{UTimeSpread} and \ref{CTimeSpread} provide strong
evidence that, quite generally, travel time spreads are largely controlled
by the same property $\omega ^{\prime }=\mathrm{d}\omega /\mathrm{d}I$ of
the background sound speed structure. Here, $2\pi /\omega (I)$ is the range
of a ray double loop and $I$ is the ray action variable. In Sec. \ref{DisCon}
our results are summarized and discussed. Two explanations for why $\omega
^{\prime }$ controls travel time spreads are given.

\section{\textsf{Background}\label{TheoBack}}

\subsection{\textsf{Theory}}

This paper is concerned with the scattering of sound, in the geometric
limit, by weak inhomogeneities. Consistent with these assumptions, our
analysis is based on the one-way form (cf. e.g. Ref.
\onlinecite{Brown-etal-03}
and references therein) of the\textit{\ ray equations},\textit{\ }
\begin{subequations}
\label{RayTra}
\begin{equation}
\frac{\mathrm{d}p}{\mathrm{d}r}=-\frac{\partial h}{\partial z},\quad \frac{%
\mathrm{d}z}{\mathrm{d}r}=\frac{\partial h}{\partial p},
\end{equation}%
and \textit{travel time equation},\textbf{\ }%
\begin{equation}
\frac{\mathrm{d}T}{\mathrm{d}r}=p\frac{\mathrm{d}z}{\mathrm{d}r}-h,
\end{equation}%
where
\end{subequations}
\begin{equation}
h(p,z,r)=-\sqrt{c^{-2}(z,r)-p^{2}}.
\end{equation}%
Here, $r$, which is the independent variable, denotes range; $z$ is depth; $%
p $ is vertical ray slowness; $T$ is travel time; and $c(z,r)$ is sound
speed. Equations (\ref{RayTra}) constitute a canonical Hamiltonian system
with $h$ the Hamiltonian, $(z,p)$ the generalized coordinate--conjugate
momentum pair, and $T$ playing the role of the mechanical action. It follows
from Eqs. (\ref{RayTra}a) and $\mathrm{d}z/\mathrm{d}r=\tan \varphi $, where
the ray angle $\varphi $ is measured relative to the horizontal, that $%
cp=\sin \varphi .$ We shall assume the sound speed to be the sum of a
range-independent background component, $C(z)$, plus a (weak)
range-dependent perturbation, $\delta c(z,r)$. This allows us to write the
Hamiltonian as the sum of a range-independent (integrable) component, $%
H(p,z)=-\sqrt{C^{-2}(z)-p^{2}}$, plus a range-dependent (nonintegrable)
perturbation, $\delta h(p,z,r)\approx -\delta c(z,r)/[C^{3}(z)H(p,z)].$

In the background environment, i.e. when $\delta c=0,$ the ray motion is
naturally described using\textit{\ action--angle variables\ }$(I,\vartheta )$%
. The transformed ray equations maintain their Hamiltonian structure with
Hamiltonian $\bar{H}(I)=H(p(I,\vartheta ),z(I,\vartheta )),$ namely
\begin{subequations}
\label{RayTra0}
\begin{equation}
\frac{\mathrm{d}I}{\mathrm{d}r}=-\frac{\partial \bar{H}}{\partial \vartheta }%
=0,\quad \frac{\mathrm{d}\vartheta }{\mathrm{d}r}=\frac{\partial \bar{H}}{%
\partial I}=\omega (I).
\end{equation}%
The travel time equation, in turn, reads%
\begin{equation}
\frac{\mathrm{d}T}{\mathrm{d}r}=I\omega -\bar{H}+\frac{\mathrm{d}}{\mathrm{d}%
r}\left( G-I\vartheta \right) .
\end{equation}%
Here,
\end{subequations}
\begin{equation}
G(z,I)=\pi I\pm \int_{z_{-}}^{z}\mathrm{d}\xi \,\sqrt{C^{-2}(\xi )-\bar{H}%
^{2}(I)},  \label{G}
\end{equation}%
where the $\pm $ sign applies to $\pm p,$ is the (multiply-valued)
generating function of the \textit{canonical transformation }that relates
implicitly the new variables\textit{\ }$(I,\vartheta )$ to the original
variables $(p,z)$ through
\begin{equation}
p=\frac{\partial G}{\partial z},\quad \vartheta =\frac{\partial G}{\partial I%
}.  \label{theta}
\end{equation}%
The action variable is defined by
\begin{equation}
I=\frac{1}{2\pi }\oint \mathrm{d}z\,p=\frac{1}{\pi }\int_{z_{-}}^{z_{+}}%
\mathrm{d}z\,\sqrt{C^{-2}(z)-H^{2}},  \label{I}
\end{equation}%
where the loop integral is taken along an isoline of $H$ and thus $z_{\pm }$
correspond to the ray upper $(+)$ and lower $(-)$ turning depths.

Equations (\ref{RayTra0}a) show that in the background environment each ray
trajectory forms a closed loop in $(p,z)$-space, or phase space, on which $I$
is a constant. The line integral in Eq. (\ref{I}) can also be expressed as
an integral over the enclosed area in phase space. It follows from Eq. (\ref%
{I}) that $I=0$ for the axial ray and that $I$ increases monotonically as
axial ray angle increases. For rays making multiple loops $G$ advances by $%
2\pi I$ and $\vartheta $ advances by $2\pi $ each time a loop is completed.
Following a ray, the difference $G-I\vartheta $ makes small oscillations
about zero. The term $\mathrm{d}(G-I\vartheta )/\mathrm{d}r$ gives small
endpoint contributions to $T.$ The fractional contributions from this term
to $T$ are nonnegligible only at very short range. The contribution from
this term to several of the analytical travel time spread estimates
presented below is zero. In those cases where this contribution is not
identically zero, it is very small.

In range-dependent environments, i.e. with $\delta c\neq 0,$ action--angle
variables can be defined using the same relations as in the unperturbed
environment \cite{Abdullaev-Zaslavsky-91}. This is in fact possible because
the range-independent canonical transformation $(p,z)\mapsto (I,\vartheta )$
depends solely on the geometry of phase space, characterized by the
symplectic two-form $\mathrm{d}p\wedge \mathrm{d}z,$ which is the same with $%
\delta c=0$ or $\delta c\neq 0;$ it does \textit{not} depend on the
Hamiltonian itself (cf. e.g. Ref.
\onlinecite{Arnold-89}%
). It is important to realize, however, that in the presence of a
range-dependent pertubation $\delta c,$ the action variable $I$ as defined
in Eq. (\ref{I}) is \textit{not} constant anymore because $H$ is \textit{not}
a first integral of (\ref{RayTra}). The Hamiltonian in the new variables
takes the form $\bar{H}(I)+\delta \bar{h}(I,\vartheta ,r)$, where $\bar{H}%
(I)=H(p(I,\vartheta ),z(I,\vartheta ))$ and $\delta \bar{h}(I,\vartheta
,r)=\delta h(p(I,\vartheta ),z(I,\vartheta ),r).$ The ray and travel time
equations then read
\begin{subequations}
\label{RayTra1}
\begin{equation}
\frac{\mathrm{d}I}{\mathrm{d}r}=-\frac{\partial \delta \bar{h}}{\partial
\vartheta },\quad \frac{\mathrm{d}\vartheta }{\mathrm{d}r}=\omega +\frac{%
\partial \delta \bar{h}}{\partial I},
\end{equation}%
and%
\begin{equation}
\frac{\mathrm{d}T}{\mathrm{d}r}=I\frac{\mathrm{d}\vartheta }{\mathrm{d}r}-%
\bar{H}-\delta \bar{h}+\frac{\mathrm{d}}{\mathrm{d}r}\left( G-I\vartheta
\right) .
\end{equation}

In realistic ocean environments Eqs. (\ref{RayTra1}) are difficult to apply
because in such environments the dependence of $\delta h$ on $I$ and $%
\vartheta $ is not known explicitly. Because of this, most of the results
presented below assume that the environment is piecewise range-independent.
These results follow from piecewise application of Eqs. (\ref{RayTra0}).
This procedure eliminates difficulties associated with the lack of
availability of an explicit expression for $\delta \bar{h}(I,\vartheta ,r).$
The endpoint correction term $\mathrm{d}(G-I\vartheta )/\mathrm{d}r$ in (\ref%
{RayTra0}b) will be neglected throughout the reminder of this paper.

The reason for introducing action--angle variables is that these variables
provide the most succinct description of the underlying ray dynamics. The
equations presented in the next three sections that make use of
action--angle variables can also be written in terms of the original phase
space variables $(p,z),$ but the $(p,z)$ form of the equations are more
complicated and obscure the important role, described below, played by $%
\omega ^{\prime }$ in controlling travel time spreads.

\subsection{\textsf{Some qualitative features of scattered ray travel time
distributions}}

Some basic features of ray travel times in deep-ocean environments without
and with internal-wave-induced scattering are shown in Fig. \ref{TimeFronts}%
. Travel time and ray depth are plotted at $r=2$ \textrm{Mm }for rays
emitted from an axial source in each of the two sound channels shown in Fig. %
\ref{SndProf}. The nonscattered ray travel times plotted in Fig. \ref%
{TimeFronts} correspond to rays with launch angles $\varphi _{0}$ confined
to $4^{\circ }$ bands. The scattered ray travel times plotted in Fig. \ref%
{TimeFronts} correspond to an ensemble of $200$ rays with a fixed launch
angle, each in the same background sound speed structure but with an
independent realization of the internal-wave-induced perturbation
superimposed. (Independent realizations were generated using the same
internal wave field by staggering the initial range at which rays were
launched with $\Delta r=5$ \textrm{km}; the vertical derivative of
internal-wave-induced sound speed perturbations has a horizontal correlation
length shorter than $5$ \textrm{km }\cite{Brown-Viechnicki-98}, so this
simple procedure ensures statistical independence.) Figure \ref{TimeFronts}
shows two examples of a portion of what is commonly referred to as a\textit{%
\ timefront}, consisting of many smooth branches that meet at cusps.

\begin{figure}[tbp]
\centerline{\includegraphics[width=8cm,clip=]{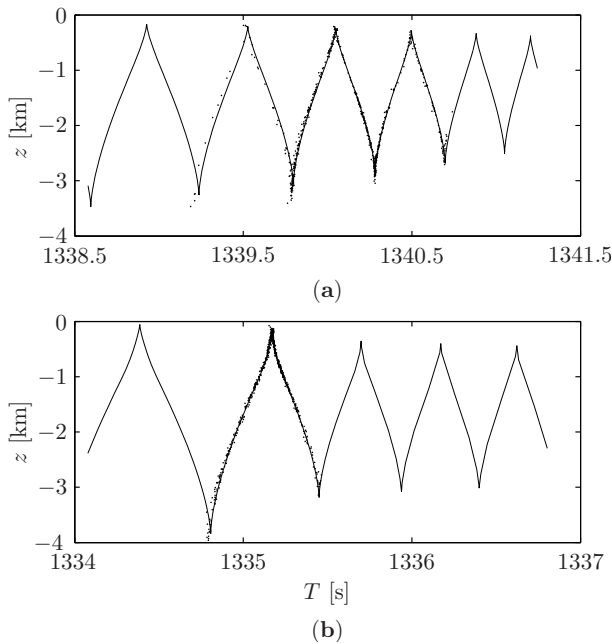}}
\caption{(a) Ray travel time and depth at $r=2$ Mm in the C0 sound channel
(cf. Fig. 2) without (solid curve) and with (dots) internal-wave-induced
sound speed perturbations superimposed. The source is on the sound channel
axis in both cases. In the absence of internal-wave-induced perturbations $%
8^{\circ }\leq \protect\varphi _{0}\leq 12^{\circ },$ where $\protect\varphi %
_{0}$ is ray launch angle. In the presence of
internal-wave-induced perturbations $\protect\varphi
_{0}=10^{\circ }$ in each of 200 realizations of the internal wave
field. (b) Same as (a) except that the C14 sound speed was used
with $10^{\circ }\leq \protect\varphi _{0}\leq 14^{\circ }$
without internal-wave-induced perturbations, and $\protect\varphi
_{0}=12^{\circ }$ with internal-wave-induced
perturbations.}\label{TimeFronts}
\end{figure}

The internal-wave-induced sound speed perturbation used to produce Fig. \ref%
{TimeFronts}---and all subsequent numerical calculations shown in this paper
that include internal-wave-induced sound speed perturbations---was computed
using Eq. (19) of Ref.
\onlinecite{Colosi-Brown-98}%
. In that expression $y$ and $t$ were set to zero, i.e. a frozen vertical
slice of an internal wave field was assumed. The range-averaged buoyancy
frequency profile measured during the AET experiment was used. The
dimensionless strength parameters $E$ and $\mu $ were taken to be $6.3\times
10^{-5}$ and $17.3,$ respectively. Horizontal wavenumber and vertical mode
number cutoffs of $2\pi $ $\mathrm{km}^{-1}$ and $30,$ respectively, were
used. The resulting sound speed perturbation field is highly structured and
fairly realistically describes a typical deep-ocean midlatitude
internal-wave-induced sound speed perturbation.

\begin{figure}[tbp]
\centerline{\includegraphics[width=8cm,clip=]{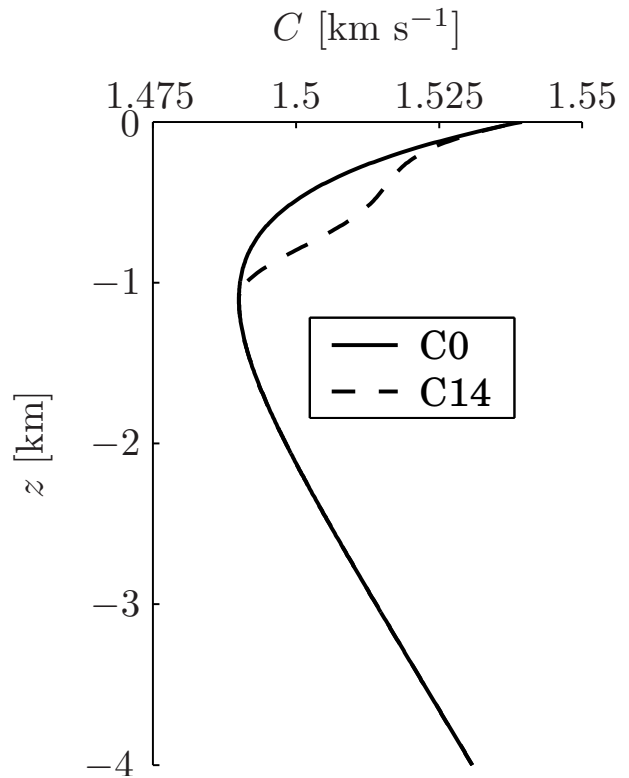}}
\caption{Background sound speed profiles used in the numerical
work presented in Figs. 1, 3, 4, 8, 9 and 10.}\label{SndProf}
\end{figure}

Figure \ref{TimeFronts} shows that internal-wave-induced scattering is
predominantly along the background timefront. Internal-wave-induced
scattering also causes a broadening of individual branches of the timefront.
This broadening occurs even for a single realization of the sound speed
perturbation field if rays with a dense set of launch angles are plotted. In
the two sections that follow, simple analytic expressions that describe time
spreads are derived and compared to numerical simulations. These expressions
describe: (i) the relatively large scattering-induced travel time spreads
along the timefront seen in Fig. \ref{TimeFronts} as a function of range,
without regard to the depth or timefront branch on which a scattered ray
falls; (ii) the relatively large scattering-induced travel time spreads of
rays whose turning history and final depth are fixed, but whose final range
is not; and (ii) the relatively small scattering-induced broadening of an
individual branch of the timefront at a fixed depth and range as a function
of range or ray double loops. We refer to (i) and (ii) as\textit{\
unconstrained travel time spreads\ }and (iii) as a\textit{\ constrained
travel time spread. }The latter is constrained by the imposition of an
eigenray constraint. Both unconstrained and constrained measures of time
spreads will be shown below to be largely controlled by $\omega ^{\prime
}(I).$

It is not surprising that time spreads in the presence of sound speed
perturbations are largely controlled by $\omega ^{\prime }$ inasmuch as
travel time dispersion in the absence of perturbations is controlled by $%
\omega ^{\prime }.$ Equation (\ref{RayTra0}b), with the term $\mathrm{d}%
\left( G-I\vartheta \right) /\mathrm{d}r$ neglected, can be integrated
immediately to give
\end{subequations}
\begin{equation}
T(I;r)=[I\omega (I)-\bar{H}(I)]r
\end{equation}%
where, from the first of Eqs. (\ref{RayTra0}a), $I$ is constant following a
ray. It follows that%
\begin{equation}
\mathrm{d}T/\mathrm{d}I=I\omega ^{\prime }r,  \label{dTdI}
\end{equation}%
where the second of Eqs. (\ref{RayTra0}a) has been used. This simple
expression succinctly describes the travel time dispersion seen in Fig. \ref%
{TimeFronts}. In both environments $\omega ^{\prime }$ is negative for all
nonaxial rays; because $I$ is a nonnegative monotonically increasing
function of $\left| \varphi _{0}\right| $ for an axial source, (\ref{dTdI})
describes the decrease in travel time with increasing $\left| \varphi
_{0}\right| $ that is shown in Fig. \ref{TimeFronts}.

\begin{figure}[tbp]
\centerline{\includegraphics[width=8cm,clip=]{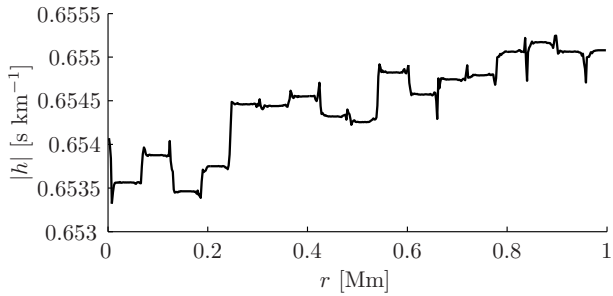}}
\caption{Hamiltonian $h$ vs. range for one of the scattered rays shown in
Fig. 1b in the presence of an internal-wave-induced sound speed perturbation
field.}
\label{HamVsRange}
\end{figure}

For one of the rays used to produce Fig. \ref{TimeFronts}, the Hamiltonian $%
h $ (whose numerical value is identical to that of $\bar{H}+\delta \bar{h}$,
although they depend on different variables) as a function of range is shown
in Fig. \ref{HamVsRange}. It is seen that $h$ vs. $r$ consists of a sequence
of approximately piecewise-constant segments, separated by fairly narrow
transition regions. The transition regions coincide with each of the rays
upper turning points. A useful and widely used approximation, the\textit{\
apex approximation}, assumes that the width of each transition region is
negligibly small. In the action--angle description of ray motion that makes
use of the apex approximation, $I$ is piecewise constant following a ray,
making jumps $\Delta I$ of negligibly small width $\Delta \vartheta $ at
each upper turning depth of the ray; between such jumps $\vartheta $
advances by $2\pi .$ A slightly relaxed form of the apex approximation in
which $\Delta \vartheta $ is treated as a small parameter will be used in
Sec. \ref{CTimeSpread}. Generally, the apex approximation works fairly well
in typical midlatitude deep ocean environments for rays with axial angles
greater than about $10\unit{%
{{}^\circ}%
}$; it is usually a poor approximation for rays with axial angles less than
approximately $5\unit{%
{{}^\circ}%
}.$ More generally, the apex approximation can be thought of as a special
case of a scattering model in which $I$ is piecewise constant on a sequence
of range intervals of variable extent. Such a model is used in Secs. \ref%
{UTimeSpread} and \ref{CTimeSpread}. Although the apex approximation is a
special case, it is useful to focus on this special case when considering
constrained spreads because it is often the case \cite%
{Duda-etal-92,Worcester-etal-99,Colosi-etal-99} that only the steep ray
travel time spreads can be measured experimentally.

\section{\textsf{Unconstrained time spreads}\label{UTimeSpread}}

\subsection{\textsf{Spreading along the timefront}}

In this section we imagine launching rays from a fixed source location with
a fixed launch angle in an ensemble of oceans, each with the same background
sound speed structure, but with an independent realization of the
internal-wave-induced perturbation field superimposed. At a fixed range we
consider the resulting distribution of ray travel time perturbations without
regard to the final ray depth or the ray turning point history. Such a
distribution is seen in Fig. \ref{TimeFronts}.

With the assumption that $I$ is\textit{\ }piecewise constant following a
ray, it follows from\ Eqs. (\ref{RayTra0}), neglecting $\mathrm{d}\left(
G-I\vartheta \right) /\mathrm{d}r$, that
\begin{equation}
\Delta T=\int_{0}^{r}\mathrm{d}\xi \,I(\xi )\omega ^{\prime }(I(\xi ))\Delta
I(\xi ).  \label{dT}
\end{equation}%
Here, $\Delta T=T-T_{0},$ where $T_{0}(r)$ is the travel time of the
unperturbed ray whose initial action is $I_{0};$ $\Delta I=I-I_{0}$; and $I$
and $\omega ^{\prime }$ are understood to be piecewise constant functions of
$r$. If the product $I\omega ^{\prime }\approx I_{0}\omega ^{\prime }(I_{0})$
over the domain of $I$-values assumed by the ray between ranges $0$ and $r,$
then (\ref{dT}) can be approximated by%
\begin{equation}
\Delta T\approx I_{0}\omega ^{\prime }(I_{0})\int_{0}^{r}\mathrm{d}\xi
\,\Delta I(\xi ).  \label{dTa}
\end{equation}%
Successive upper turns are typically separated by about $50$ \textrm{km},
which is large compared to the horizontal correlation length of
internal-wave-induced sound speed fluctuations, so each perturbation to $I$
is independent. This leads to $\langle (\Delta I)^{2}\rangle ^{1/2}\sim
r^{1/2}$, where $\langle \cdot \rangle $ indicates ensemble average, and, in
turn, to $\langle (\Delta T)^{2}\rangle ^{1/2}\sim r^{3/2}.$ This $r^{3/2}$%
-dependence was previously derived by F. Henyey and J. Colosi (personal
communication), who also found that this dependence is in good agreement
with simulations. [Our simulations also indicate that when (\ref{dTa}) is
valid, then $\langle (\Delta T)^{2}\rangle ^{1/2}\sim r^{3/2};$ our focus,
however, is on the dependence of (\ref{dT}) and (\ref{RayTra1}) on the
background sound speed structure, via $I\omega ^{\prime }$. We note, in
addition, that if $\langle (\Delta I)^{2}\rangle ^{1/2}$ is computed using
an ensemble of rays with the same launch angle then the $r$-dependence of
this quantity is characterized by oscillations in $r$ superimposed on the $%
r^{3/2}$ trend. The oscillations are caused by the cycling between
relatively small time spreads when distributions are centered near the cusps
where neighboring timefront branches join and relatively large time spreads
when distributions are centered near the midpoint of timefront branches. At
long range when scattered ray distributions are very broad, this effect is
greatly dimished.] At long range the contributions to (\ref{dT}) or (\ref%
{dTa}) from ray partial loops near the source and receiver are unimportant.
Then, because $2\pi /\omega $ is the range of a ray double loop, for a ray
that has undergone $N$ apex scattering events and has $N$ complete double
loops, (\ref{dTa}) can be further approximated,%
\begin{equation}
\Delta T\approx 2\pi \frac{I_{0}\omega ^{\prime }(I_{0})}{\omega (I_{0})}%
\sum_{i=1}^{N}(I_{i}-I_{0}).  \label{dTal}
\end{equation}

Equation (\ref{dTal}) provides an explanation for the observation (recall
Fig. \ref{TimeFronts}) that $\varphi _{0}=10^{\circ }$ rays in the C0
environment are more strongly scattered than $\varphi _{0}=12^{\circ }$ rays
in the C14 environment; the same internal wave field was used to generate
the ensemble of scattered rays, so the difference between these figures is
due to the difference in the background sound speed profiles. The $\varphi
_{0}=12^{\circ }$ rays in the C14 environment have a $|\omega ^{\prime
}(I_{0})|$-value that is approximately one-fourth that of the $\varphi
_{0}=10^{\circ }$ rays in the C0 environment, and this scattering-induced
time spread is seen to be reduced by a nearly commensurate amount,
consistent with Eq. (\ref{dTal}). According to that equation, travel time
perturbations are the product of the amplification factor $I\omega ^{\prime
}/\omega $ and a term that depends on the history of the scattering-induced
perturbations to the ray action variable.

\begin{figure}[tbp]
\centerline{\includegraphics[width=8cm,clip=]{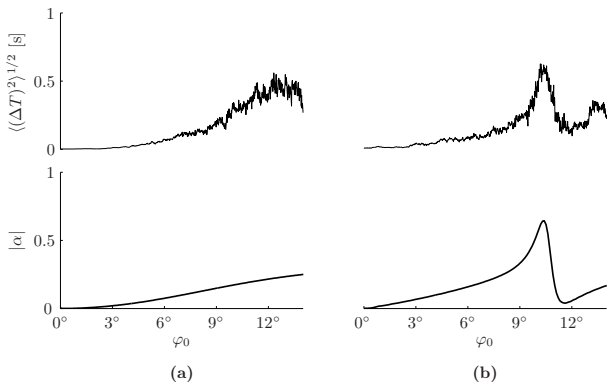}}
\caption{Estimates of unconstrained travel time spreads (upper panels) and
the stability parameter $\protect\alpha $ (lower panels) as a function of
launch angle in background sound channels C0 (a) and C14 (b) (cf. Fig. 2).
To generate the upper plots, ray simulations based on Eqs. (1) were
performed by launching rays from a source located at the sound channel axis
with a fixed launch angle in an ensemble of 50 oceans, each with the same
background sound speed structure, but with an independent realization of the
internal-wave-induced perturbation field superimposed.}
\label{UDeltaT}
\end{figure}

It is convenient to introduce the\textbf{\ }\textit{stability parameter}%
\textbf{\ }%
\cite{Abdullaev-Zaslavsky-91,Smirnov-Virovlyansky-Zaslavsky-01,Beron-Brown-03}%
\begin{equation}
\fbox{$\alpha =I\omega ^{\prime }/\omega ,$}
\end{equation}%
which, in addition to appearing in Eq. (\ref{dTal}), is a natural
nondimensional measure of $\omega ^{\prime }.$ Figure \ref{UDeltaT} compares
numerically simulated time spreads $\langle (\Delta T)^{2}\rangle ^{1/2}$ as
a function of launch angle $\varphi _{0}$ for a source on the sound channel
axis at $r=2$ $\mathrm{Mm}$ to $\alpha $ in the two different background
environments shown in Fig. \ref{SndProf} on which identical
internal-wave-induced sound speed perturbation fields were superimposed. In
the Fig. \ref{UDeltaT} plots, dependence on $I_{0}$ is replaced by
dependence on the more familiar variable $\varphi _{0}$; for an axial source
this constitutes a simple stretching of the horizontal axis. Ensembles of $%
\Delta T$ for rays with the same launch angle were generated using the same
technique that was used to generate Fig. \ref{TimeFronts}.

Figure \ref{UDeltaT} shows that in both environments almost all of the
structure seen in $\langle (\Delta T)^{2}\rangle ^{1/2}$ can be attributed
to the stability parameter $\alpha .$ This is because in the environments
considered Eq. (\ref{dTal}) is generally a good approximation to Eq. (\ref%
{dT}). Also, the dependence of $\sum_{i=1}^{N}(I_{i}-I_{0})$ on axial ray
angle is somewhat weaker than the dependence of $\alpha $ on axial ray
angle. We conclude from Fig. \ref{UDeltaT} that travel time spreads along
the timefront are largely controlled by the background sound speed structure
via $\alpha $, and that, provided $\alpha (I)$ does not have too much
structure, the dependence of $\Delta T$ on $\alpha $ in Eq. (\ref{dT}) can
be taken outside the integral.

\begin{figure}[tbp]
\centerline{\includegraphics[width=8cm,clip=]{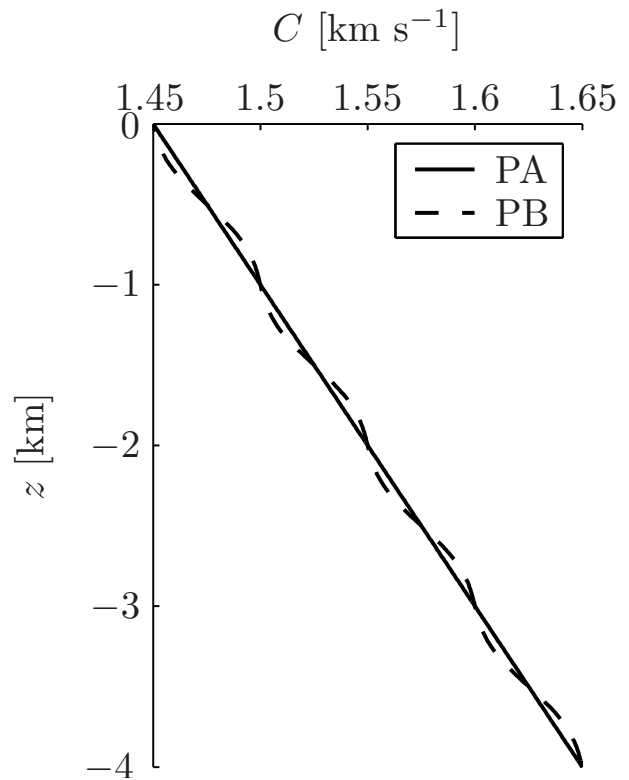}}
\caption{Background sound speed profiles used to construct the curves in
Fig. 6.}
\label{SndProfPolar}
\end{figure}

\subsection{\textsf{Spreading of rays with fixed turning history and final
depth}}

We now consider a different measure of unconstrained travel time spread. To
illustrate the generality of the results presented, we consider rays in
upward refracting environments (Fig. \ref{SndProfPolar}) in the presence of
rough surface scattering. In such environments the action--angle form of the
ray equations is unchanged. The action $I$ is defined as in (\ref{I}) with
the upper turning depth $z_{+}=0$ for all rays. It follows from Eqs. (\ref%
{RayTra0}), neglecting $\mathrm{d}\left( G-I\vartheta \right) /\mathrm{d}r$,
that the travel time for one ray cycle is $T(I)=2\pi \lbrack I-\bar{H}%
(I)/\omega (I)].$ At each reflection from the rough surface the ray action
gets modified. After $N$ surface reflections and $N+1$ ray cycles the travel
time perturbation to the ray is%
\begin{equation}
\Delta T_{N}\approx 2\pi \frac{\bar{H}(I_{0})\omega ^{\prime }(I_{0})}{%
[\omega (I_{0})]^{2}}\sum_{i=1}^{N}(I_{i}-I_{0}),  \label{dTrs}
\end{equation}%
where $I_{0}$ is the initial action of the ray, and the first of Eqs. (\ref%
{RayTra0}a) has been used. Note that Eq. (\ref{dTrs}) constrains the ray
geometry and final ray depth but not the final range of the scattered rays.

\begin{figure}[tbp]
\centerline{\includegraphics[width=8cm,clip=]{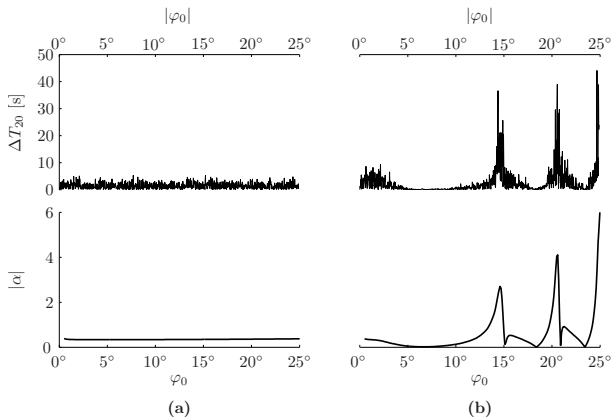}}
\caption{Upper panels: absolute value of the difference between perturbed
(rough surface) and unperturbed (flat surface) ray travel time after 20
surface reflections as a function of initial ray angle at the surface for
sound speed profiles PA (a) and PB (b) shown in Fig. 5. Lower panels:
stability parameter as a function of ray angle in each of the two
environments.}
\label{UDeltaTrs}
\end{figure}

Figure \ref{UDeltaTrs} shows plots of $\Delta T_{20},$ defined in (\ref{dTrs}%
), vs. launch angle $\varphi _{0}$ for a source at the surface [so, again, $%
\varphi _{0}(I_{0})$ is a simple stretching] in the environments shown in
Fig. \ref{SndProfPolar}. The rough surface was a single frozen realization
of a surface gravity wavefield with a $k^{-7/2}$ surface elevation
wavenumber spectrum with $0.02$ rad m$^{-1}\leq k\leq 0.16$ rad m$^{-1},$ $%
\Delta k=10^{-3}$ rad m$^{-1},$ and an r.m.s. slope of $4\times 10^{-3}.$
Ray reflections from this surface were specular, but with a linearized
boundary condition; the surface elevation was neglected, but the nonzero
slope was not approximated. The same environments were used in Ref.
\onlinecite{Beron-Brown-03}%
.

Consistent with Eq. (\ref{dTrs}), Fig. \ref{UDeltaTrs} shows that travel
time perturbations are largely controlled by the background sound speed
structure via $\alpha $, rather than details of the rough surface.

\section{\textsf{Constrained time spread}\label{CTimeSpread}}

In this section we focus on deep ocean conditions and consider the
scattering-induced broadening of an individual branch of the timefront. Note
that this broadening is much smaller than the scattering-induced spreading
along the timefront that was considered in section \ref{UTimeSpread}.A. We
refer to the broadening of an individual branch of the timefront as a
constrained time spread because to calculate this broadening two constraints
must be incorporated into the calculation. First, the rays contributing to
the spread must have the same fixed endpoints in the $(z,r)$-plane. Second,
the rays contributing to the spread must have the same turning point
history, i.e. the same ray inclination (positive or negative) at the source,
and the same number of turning points $M$ (upper and lower) between source
and receiver. Note, however, that these constraints do not fix the values of
the ray angle at either the source or receiver.

The approach taken here to compute appropriately constrained travel time
spreads is based on a perturbation expansion that exploits the assumed
smallness of the sound speed perturbation $\delta c$. The method by which
the constraints are incorporated into the travel time perturbation estimates
presented below is different than the method that was used in Ref.
\onlinecite{Beron-etal-03}%
; that approach can be shown to give the same result that is presented below
when terms of $O((\Delta I)^{2})$, neglecting endpoint corrections, are
retained.

The perturbation expansions presented below make use of a scattering model
in which $I$ following a scattered ray is piecewise constant; the apex
approximation is then treated as a special case. Away form the scattering
events Eqs. (\ref{RayTra0}) are valid piecewise, with $I$ making a jump at
each scattering event. Thus Eqs. (\ref{RayTra0}) can be used to compute, in
a piecewise fashion, the contribution of each ray segment to the travel time
and range of the scattered ray. To impose the eigenray constraint, the total
range of each scattered ray must be constrained to be equal to the range of
the unperturbed ray. In addition, the turning point history, starting depth,
and ending depth of the scattered ray must be the same as those of the
unperturbed ray. This is accomplished by constraining the total change in $%
\vartheta $ following the scattered ray to be equal to that of the
unperturbed ray. For some choices of the source--receiver geometry this
procedure exactly enforces the eigenray constraint. In general, however,
small endpoint corrections must be applied. For simplicity, these small
endpoint corrections, which are important only at very short range, will be
neglected.

We assume that between source and receiver the environment can be
approximated as $n$ piecewice range-independent segments. Let $I_{i}$ denote
the action of a scattered ray in the $i$th segment whose horizontal extent
is $\Delta r_{i}=\Delta \vartheta _{i}/\omega (I_{i}),$ and let $\tilde{I}$
denote the action of the unperturbed ray. The eigenray constraint requires
the perturbed ray total range, $r$, to be equal to the unperturbed ray total
range. This condition reads%
\begin{equation}
r=\sum_{i=1}^{n}\frac{\Delta \vartheta _{i}}{\omega (I_{i})}=\frac{1}{\omega
(\tilde{I})}\sum_{i=1}^{n}\Delta \vartheta _{i},  \label{eigc}
\end{equation}%
which, for sufficiently small $\Delta I_{i}=I_{i}-\tilde{I}$, can be
approximated by%
\begin{equation}
\frac{\omega ^{\prime }(\tilde{I})}{[\omega (\tilde{I})]^{2}}%
\sum_{i=1}^{n}\Delta I_{i}\Delta \vartheta _{i}=\frac{1}{2}\left. \frac{%
\mathrm{d}^{2}}{\mathrm{d}I^{2}}\right| _{I=\tilde{I}}\left( \frac{1}{\omega
}\right) \sum_{i=1}^{n}(\Delta I_{i})^{2}\Delta \vartheta _{i}.  \label{EigC}
\end{equation}%
The r.h.s. term of (\ref{EigC}) is almost always negligible compared to the
l.h.s. term in deep ocean environments, even at ranges comparable to basin
scales. Then, because $\Delta \vartheta _{i}=\omega (I_{i})\Delta
r_{i}\approx \omega (\tilde{I})\Delta r_{i},$ the eigenray constraint
reduces approximately to a statement that the range-weighted average of the
perturbed action is equal to the unperturbed action, i.e.%
\begin{equation}
\tilde{I}\approx r^{-1}\sum_{i=1}^{n}I_{i}\Delta r_{i}.  \label{EigI}
\end{equation}%
This equation constrains the action history of scattered eigenrays.

The difference between the perturbed and unperturbed ray travel times is
\begin{eqnarray}
\Delta T &=&\sum_{i=1}^{n}\left[ I_{i}-\frac{\bar{H}(I_{i})}{\omega (I_{i})}%
\right] \Delta \vartheta _{i}-\left[ \tilde{I}-\frac{\bar{H}(\tilde{I})}{%
\omega (\tilde{I})}\right] \sum_{i=1}^{n}\Delta \vartheta _{i}  \notag \\
&\approx &\frac{\omega ^{\prime }(\tilde{I})\bar{H}(\tilde{I})}{[\omega (%
\tilde{I})]^{2}}\sum_{i=1}^{n}\Delta I_{i}\Delta \vartheta _{i}-\frac{1}{2}%
\left. \frac{\mathrm{d}^{2}}{\mathrm{d}I^{2}}\right| _{I=\tilde{I}}\left(
\frac{\bar{H}}{\omega }\right) \sum_{i=1}^{n}(\Delta I_{i})^{2}\Delta
\vartheta _{i}  \notag \\
&=&\frac{1}{2}\frac{\omega ^{\prime }(\tilde{I})}{\omega (\tilde{I})}%
\sum_{i=1}^{n}(\Delta I_{i})^{2}\Delta \vartheta _{i},  \label{CdT}
\end{eqnarray}%
where the eigenray constraint (\ref{EigC}) has been used. Virovlyansky \cite%
{Virovlansky-03} recently derived an expression for $\Delta T$ consisting of
Eq. (\ref{CdT}) [written as an integral over $r$ using $\Delta \vartheta
_{i}=\omega (I_{i})\Delta r_{i}$] plus a correction term. We will discuss
his result in more detail below. It should also be noted that the same
symbol $\Delta T$ is being used to denote a constrained (eigenray) travel
time perturbation that was used in the previous section to denote an
unconstrained travel time perturbation. We believe that is obvious in all
cases which is the relevant quantity. This choice was made to keep the
notation simple. Also, for the same reason, we make no notational
distinction between a theoretical estimate of $\Delta T$ and a numerically
computed $\Delta T.$

The apex approximation is a special case of the analysis leading to Eqs. (%
\ref{eigc})-(\ref{CdT}). If the contributions to $\Delta T$ from the
incomplete ray cycles at the beginning and end of the ray path are
neglected, Eq. (\ref{CdT}) applies with $\Delta \vartheta _{i}=2\pi $ and $%
n=N,$ the number of complete ray cycles. (For large $N$ the neglected
incomplete ray cycle contributions to $\Delta T$ constitute small
corrections to the sum of the retained contributions.)

In Ref.
\onlinecite{Beron-etal-03}
a relaxed form of the apex approximation was considered. The transition
width of the jump $\Delta I_{\mathrm{T}}=I_{i}-I_{i-1}$ had a width $\Delta
\vartheta _{\mathrm{T}},$ which is assumed here to be small, $\Delta
\vartheta _{\mathrm{T}}<2\pi .$ In the transition region a particular form $%
\bar{h}(I-\vartheta \Delta I_{\mathrm{T}}/\Delta \vartheta _{\mathrm{T}})$
of the Hamiltonian was assumed. This finite width transition region was
shown not to contribute to a range perturbation, but gives a travel time
perturbation $\frac{1}{2}\Delta I_{\mathrm{T}}\Delta \vartheta _{\mathrm{T}}$
for a single scattering event. After $N$ scattering events, assuming the
transition width $\Delta \vartheta _{\mathrm{T}}$ of each is the same, this
gives a contribution $\frac{1}{2}(I_{N}-I_{0})\Delta \vartheta _{\mathrm{T}}$
to $\Delta T$ (\ref{CdT}), while the eigenray constraint (\ref{EigC}) is
unaltered. With this additional term (\ref{CdT}), with $\Delta \vartheta
_{i}=2\pi $ and $n=N,$ is replaced by
\begin{equation}
\Delta T=\dfrac{1}{2}(I_{N}-I_{0})\Delta \vartheta _{\mathrm{T}}+\pi \frac{%
\omega ^{\prime }(\tilde{I})}{\omega (\tilde{I})}\sum_{i=1}^{N}(\Delta
I_{i})^{2}.  \label{CdTr}
\end{equation}%
For convenience we shall refer to the first and second terms on the r.h.s.
of Eq. (\ref{CdTr}) as $\Delta T_{1}$ and $\Delta T_{2},$ respectively. For
steep rays in typical deep ocean environments $\Delta \vartheta _{\mathrm{T}%
}/(2\pi )\approx 0.05$ while $\omega ^{\prime }\Delta I_{\mathrm{T}}/\omega
\approx 0.01.$ (The latter estimate is variable owing to variations in $%
\omega ^{\prime }.$) Thus one might expect that $\Delta T_{1}$ dominates $%
\Delta T_{2}$. For small $N$ this is indeed the case. But under typical deep
ocean conditions consecutive apex scattering events are independent (because
the ray cycle distance $2\pi /\omega \approx 50$ \textrm{km} exceeds the
horizontal correlation length of internal waves, about $10$ \textrm{km}), so
$\Delta T_{1}$ and $\Delta T_{2}$ grow approximately like $N^{1/2}$ and $N,$
respectively. Thus $\Delta T_{2}$ is expected to dominate $\Delta T_{1}$ at
long range. Note that when $\Delta T_{2}$ dominates $\Delta T_{1}$, this
equation predicts a scattering-induced travel time bias in the direction of $%
\limfunc{sgn}\omega ^{\prime }.$%
\begin{figure}[tbp]
\centerline{\includegraphics[width=8cm,clip=]{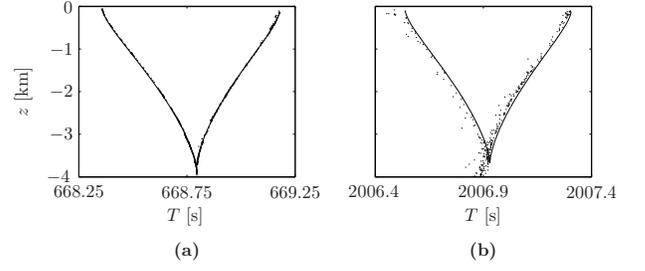}}
\caption{(a) Ray travel time and depth of rays with positive launch angles
and which have $N=18$ upper turning points at $r=1$ \textrm{Mm }in the C0
environment shown in Fig. 2. The solid curve correponds to rays in the
absence of internal-wave-induced scattering. Dots correspond to rays
scattered by internal waves. (b) Same as the left panel except that $r=3$
\textrm{Mm} and $N=54.$}
\label{TTbias}
\end{figure}

These observations are consistent with the simulations shown in Fig. \ref%
{TTbias}. There, scattered and unperturbed ($\delta c=0$) ray travel times
are shown in the vicinity of two branches of the timefront at $r=1$ $\mathrm{%
Mm}$ ($N=18$) and $r=3$ $\mathrm{Mm}$ ($N=54$). At both ranges the source
was on the sound channel axis in the C0 environment shown in Fig. \ref%
{SndProf}; launch angles for the rays shown are near $10^{\circ }$ at both
ranges. In that environment $\limfunc{sgn}\omega ^{\prime }=-1.$ It is seen
in this figure that there is no indication of a negative travel time bias at
$1$ $\mathrm{Mm}$, while there is a clear negative bias at $3$ $\mathrm{Mm}$%
, consistent with Eq. (\ref{CdTr}). At ranges sufficiently short that $%
\Delta T_{1}$ dominates, travel time biases of either sign can be expected
as $I_{N}-I_{0}$ can be of either sign. Because $I_{N}-I_{0}$ is a
continuous function of launch angle, such biases should have a nonzero
correlation scale along the timefront. These features are readily evident in
our simulations in the presence of realistic internal-wave-induced sound
speed perturbations.

\begin{figure}[tbp]
\centerline{\includegraphics[width=8cm,clip=]{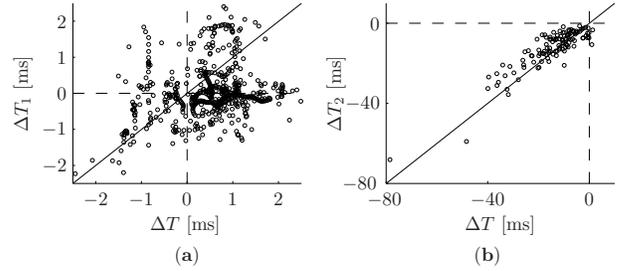}}
\caption{(a) Constrained travel time spread ($\Delta T$) vs. the first term
on the r.h.s. of Eq. (\ref{CdTr}) ($\Delta T_{1}$) at $r=500$ km for rays
with axial launch angles $11^{\circ }\leq \protect\varphi _{0}\leq
13.5^{\circ }$ in the C0 environment with an internal-wave-induced sound
speed perturbation superimposed. (b) As in (a) but vs. the second term on
the r.h.s. of Eq. (\ref{CdTr}) ($\Delta T_{2}$) and at $r=3$ Mm.}
\label{CDeltaTApex}
\end{figure}

Quantitative tests of the correctness of Eq. (\ref{CdTr}) are shown in Fig. %
\ref{CDeltaTApex}. There numerically computed travel time differences $%
\Delta T$ are compared separately to $\Delta T_{1}$ and $\Delta T_{2}$ under
conditions in which one of the two terms dominates the other. The
constrained travel time difference $\Delta T$ was computed using a single
realization of an internal-wave-induced sound speed perturbation as the
difference between the perturbed ray with the same turning history and the
same final depth as the unperturbed ray. In Fig. \ref{CDeltaTApex} attention
is restricted to rays that are sufficiently steep that the apex
approximation is approximately valid, but not so steep that rays reflect off
the surface.

In Fig. \ref{CDeltaTApex}a $\Delta T$ is compared to $\Delta T_{1}$ at $%
r=500 $ \textrm{km}; for the rays used to construct this plot the mean value
of $\Delta T_{2}$ is $-0.3$ \textrm{ms}, which is seen to represent a small
correction, on average, to $\Delta T_{1}.$ The value of $\Delta \vartheta _{%
\mathrm{T}}$ used to construct this plot is $0.25\approx 2\pi /25.$ The
domain and gross distribution of $\Delta T$ points plotted in Fig. \ref%
{CDeltaTApex}a is seen to coincide with the domain and gross distribution of
$\Delta T_{1}$ points plotted. Thus $\Delta T_{1}$ is a fairly good
statistical descriptor of $\Delta T$. The point-by-point comparison between $%
\Delta T$ and $\Delta T_{1}$ is not good, however. This is seen by noting
that the points plotted do not cluster along the diagonal line with slope
unity. Thus, although $\Delta T_{1}$ has the correct qualitative features
and appears to be a useful statistical descriptor of $\Delta T,$ it is
evidently a poor deterministic predictor of $\Delta T.$ This shortcoming, we
believe is attributable to the overly idealized form of the Hamiltonian in
the apex transition region that was used to derive $\Delta T_{1}.$ It is
interesting to note that $\Delta T_{1}$ is the only term presented in this
paper describing a travel time spread that is independent of $\omega
^{\prime },$ and this term is the one that gives the poorest agreement with
simulations.

In Fig. \ref{CDeltaTApex}b $\Delta T$ is compared to $\Delta T_{2}$ at $r=3$
\textrm{Mm}; for the rays used to construct this plot the r.m.s. value of $%
\Delta T_{1}$ is $2$ \textrm{ms}, which is seen to represent a small
correction, on average, to $\Delta T_{2}.$ The agreement between $\Delta T$
and $\Delta T_{2}$ in this plot is seen to be very good, indicating that for
the rays used to construct this plot $\Delta T_{2}$ is a very good predictor
of $\Delta T.$ Overall, for steep rays we have found good qualitative
agreement at short range between simulations of $\Delta T$ and $\Delta T_{1}$%
, and a good quantitative agreement at long range between $\Delta T$ and $%
\Delta T_{2}.$

\begin{figure}[tbp]
\centerline{\includegraphics[width=8cm,clip=]{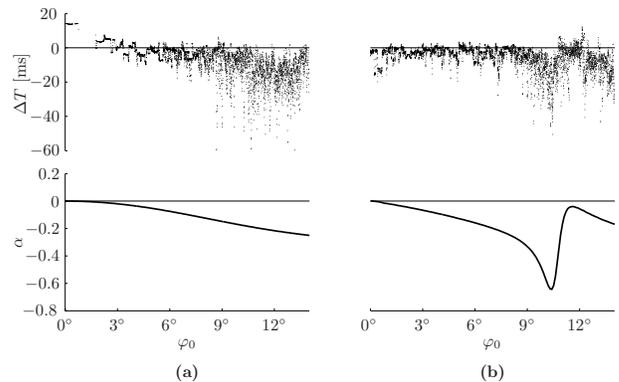}}
\caption{Travel time difference $\Delta T=T_{\mathrm{IW}}-T_{0}$ at $r=3$
\textrm{Mm} vs. ray launch angle $\protect\varphi _{0}$ for an axial source,
and stability parameter $\protect\alpha $ vs. $\protect\varphi _{0},$ in
each of the environments C0 (a) and C14 (b) shown in Fig. 2. Here, $T_{%
\mathrm{IW}}$ is the travel time of a ray in the presence of an
internal-wave-induced sound speed perturbation field, and $T_{0}$ is the
travel of a ray in the same background environment, but without the sound
speed perturbation field superimposed, whose turning point history and final
depth are the same as those of the $T_{\mathrm{IW}}$ ray.}
\label{CDeltaT}
\end{figure}

Figure \ref{CDeltaT} shows plots of $\Delta T$ vs. launch angle and $\alpha $
vs. launch angle at $r=3$ \textrm{Mm} in each of the two background sound
speed profiles shown in Fig. \ref{SndProf}. Again, the constrained travel
time spread $\Delta T$ was computed using a single realization of an
internal-wave-induced sound speed perturbation as the difference between the
perturbed ray travel time and the travel time of the unperturbed ray with
the same turning history and the same final depth as the perturbed ray. The
small gaps in the plot correspond to perturbed rays whose final depth lies
outside the bounds of the portion of the timefront which has the same
turning history as the perturbed ray. In Fig. \ref{CDeltaT} ray angles are
not limited to the band for which the apex approximation is expected to be
valid. Thus Eq. (\ref{CdTr}) is not expected to be valid for the entire band
of angles. Equation (\ref{CdT}) should be approximately valid across the
entire band, however, so the trend in $\alpha (\varphi _{0})$ should be
approximately reproduced in the $\Delta T(\varphi _{0})$ points plotted.
This is seen to be the case; variations in $\Delta T$ are caused by
variations in $\sum_{i=1}^{n}(\Delta I_{i})^{2}\Delta \vartheta _{i}$. The
probable cause of the positive values of $\Delta T$ for near-axial rays in
the C0 environment seen in Fig. (\ref{CDeltaT}) will be discussed below.
Even after fitting a smooth curve through the fluctuations seen in the Fig. %
\ref{CDeltaT} $\Delta T$ plot for the C14 environment, the peak in $|\Delta
T|$ is seen to be less pronounced than the peak in $|\alpha |.$ This is
expected because the scattering process leads to a local average over a band
of adjacent $\alpha $-values. With these minor caveats, Fig. \ref{CDeltaT}
shows that constrained travel time spreads at long range are largely
controlled by the background sound speed structure via the stability
parameter $\alpha .$ The same conclusion can be drawn from Fig. \ref%
{CDeltaTApex}b.

The probable cause of the positive computed values of $\Delta T$ for
near-axial rays seen in Fig. \ref{CDeltaT} in the C0 environment is not the
failure of Eq. (\ref{CdTr}); rather it is the failure of our ray labeling
scheme for near axial rays. We have implicitly assumed that the ray
identifier $\pm M$ uniquely identifies each timefront branch. This
assumption fails for near-axial rays. (The assumption also fails for steep
rays in the vicinity of the cusps where adjacent timefront branches join,
and whenever $\omega ^{\prime }$ has isolated zeros. We have intentionally
avoided the latter situation.) The correct way to identify timefront
branches is by the Maslov index \cite{Brown-etal-03} $\mu $ which, for waves
in two space dimensions, advances by one unit each time a caustic is
touched. For near-axial rays the portion of the timefront corresponding to a
constant value of the ray identifier consists of two adjacent partial
branches joined at a cusp. For these rays a ray-identifier-based definition
of $\Delta T$ can lead to either undefined or nonuniquely defined $\Delta T$
values.

We noted earlier the quantitative correctness of $\Delta T_{2}$ and the
qualitative correctness of $\Delta T_{1}.$ The weakest link in our analysis
is clearly $\Delta T_{1}.$ Virovlyansky \cite{Virovlansky-03} derived an
alternate correction to $\Delta T_{2}$ whose validity, like (\ref{CdT}), is
not linked to the apex approximation. In our notation, this correction reads%
\begin{eqnarray}
\Delta T_{1}^{\mathrm{V}} &=&\int_{0}^{r}\mathrm{d}\xi \,\Delta I(\xi )\frac{%
\partial \delta \bar{h}}{\partial I}(I(\xi ),\vartheta (\xi ),\xi )  \notag
\\
&&-\int_{0}^{r}\mathrm{d}\xi \,\delta h(p(\xi ),z(\xi ),\xi ).  \label{DT1V}
\end{eqnarray}%
Unfortunately, the lack of explicit knowledge on the functional dependence
of $\delta \bar{h}$ on $(I,\vartheta ,r)$ makes the numerical evaluation of
the first term on r.h.s. of (\ref{DT1V}) quite difficult---if not
impossible. Virovlyansky argued that that term can be neglected. We have
evaluated numerically the second term on r.h.s. of (\ref{DT1V}); this
resulted in significantly poorer agreement with our simulations that was
found using our $\Delta T_{1}$. (Note, however, that the parameter $\Delta
\vartheta _{\mathrm{T}}$ in our expression for $\Delta T_{1}$ was adjusted
to approximately match the simulations.) We expect, but have not confirmed,
that the evaluation of both terms on the r.h.s. of (\ref{DT1V}) would result
in good agreement with simulations.

\section{\textsf{Discussion and conclusions}\label{DisCon}}

In this paper we have investigated three measures of travel time spreads for
sound propagation in environments consisting of a range independent
background on which a range-dependent perturbation is superimposed: (i)
unconstrained spread of ray travel time along the timefront; (ii)
unconstrained spread of rays whose turning history and final depth are fixed
but whose final range is not; and (iii) scattering-induced broadening of an\
individual branch of the timefront at a fixed location. All three measures
of time spreads were shown to be largely controlled by a property $\omega
^{\prime }$ of the background sound speed profile. Surprisingly, this is the
same property that controls ray spreading and, hence, ray amplitudes \cite%
{Beron-Brown-03}. We now present two arguments that provide some insight
into why travel time spreads should be controlled by the same property of
the background sound speed profile that controls ray spreading.

The variational equations that describe how small perturbations $(\delta
I,\delta \vartheta ,\delta T)$ evolve in the extended phase space $%
(I,\vartheta ,T)$ are given by%
\renewcommand{\arraystretch}{0.75}%
\begin{equation}
\frac{\mathrm{d}}{\mathrm{d}r}\left(
\begin{array}{c}
\delta I \\
\delta \vartheta  \\
\delta T%
\end{array}%
\right) =\left[
\begin{array}{ccc}
-\delta \bar{h}_{,I\vartheta } & \delta \bar{h}_{,\vartheta \vartheta } & 0
\\
\omega ^{\prime }+\delta \bar{h}_{,II} & \delta \bar{h}_{,I\vartheta } & 0
\\
I\omega ^{\prime }+I\delta \bar{h}_{,II} & I\delta \bar{h}_{,I\vartheta
}-\delta \bar{h}_{,\vartheta } & 0%
\end{array}%
\right] \left(
\begin{array}{c}
\delta I \\
\delta \vartheta  \\
\delta T%
\end{array}%
\right) ,  \label{RayTra2}
\end{equation}%
\renewcommand{\arraystretch}{1.00}%
where a short-hand notation for partial differentiation has been introduced.
In general Eqs. (\ref{RayTra2}) and (\ref{RayTra1}) constitute a set of six
coupled differential equations. For the special case $\delta \bar{h}=0,$ $I$
and $\omega ^{\prime }$ are constant following trajectories and (\ref%
{RayTra2}), like (\ref{RayTra1}), have a simple analytical
solution. For the class of problems treated in this paper the
nonzero sound speed perturbation terms in the matrix on the r.h.s.
of (\ref{RayTra2}) are generally much smaller than the $\omega
^{\prime }$ and $I\omega ^{\prime }$ terms, corresponding to
contributions from the background sound speed profile. Thus one
expects that generically the dominant cause of the growth of
$(\delta I,\delta \vartheta ,\delta T)$ is the background sound
speed structure via $\omega ^{\prime },$ rather than the small
sound speed perturbation terms. Loosely speaking, the perturbation
terms provide a seed for the growth of $(\delta I,\delta \vartheta
,\delta T),$ while subsequently growth of these quantities is
largely controlled by $\omega ^{\prime }.$

Additional insight into the role played by $\omega ^{\prime }$ is obtained
by making a fluid mechanical analogy. Equations (\ref{RayTra}) define a flow
in the three-dimensional space $(p,z,T)$ with velocity components $\mathrm{d}%
(p,z,T)/\mathrm{d}r.$ (Recall that $r$ plays the role of the independent or
time-like variable in the one-way ray equations.) Alternatively, the flow in
this three-dimensional space can be described using action--angle variables $%
(I,\vartheta ,T).$ The coordinates $(I,\vartheta ,T)$ behave qualitatively
like cylindrical coordinates $(\rho ,\theta ,\zeta )$, say. If we neglect
the contributions from the sound speed perturbation term $\delta \bar{h},$
then the velocity components of the background flow in these coordinates are
$u_{I}=\mathrm{d}I/\mathrm{d}r=0,$ $u_{\vartheta }=(\mathrm{d}\vartheta /%
\mathrm{d}r)I=\omega I,$ and $u_{T}=\mathrm{d}T/\mathrm{d}r=I\omega -\bar{H}%
. $ The strain rate tensor,%
\begin{equation}
\mathsf{S}_{j}^{i}:=\tfrac{1}{2}\left( \nabla _{j}u^{i}+\nabla
_{i}u^{j}\right) ,
\end{equation}%
where $\nabla _{i}$ denotes covariant derivative, describes how small
elements of fluid are deformed by the flow \cite{Batchelor-64}. With the
identification $(I,\vartheta ,T)\leftrightarrow (\rho ,\theta ,\zeta )$ and
with the velocity field $(u_{I},u_{\vartheta },u_{T})$ defined above, the
strain rate tensor is%
\renewcommand{\arraystretch}{0.75}%
\begin{equation}
\mathsf{S}=\left[
\begin{array}{ccc}
0 & \frac{1}{2}I\omega ^{\prime } & \frac{1}{2}I\omega ^{\prime } \\
\frac{1}{2}I\omega ^{\prime } & 0 & 0 \\
\frac{1}{2}I\omega ^{\prime } & 0 & 0%
\end{array}%
\right] .  \label{S2}
\end{equation}%
\renewcommand{\arraystretch}{1.00}%
Although (\ref{S2}) is only qualitatively correct [because of the
qualitative connection between $(I,\vartheta ,T)$ and $(\rho ,\theta ,\zeta
) $], the conclusion to be drawn from this tensor is extremely important:
deformation of small elements of the extended phase space $(I,\vartheta ,T)$%
\ by the background sound speed structure is caused entirely by shear
(off-diagonal elements of $\mathsf{S}$)\ and is quantified by the product $%
I\omega ^{\prime }.$ This behavior is consistent with the discussion above
concerning Eq. (\ref{RayTra2}) and the growth of small perturbations in the
extended phase space $(I,\vartheta ,T).$

Arguments similar to those given above relating to Eqs. (\ref{RayTra2}) and (%
\ref{S2}) were given in Ref.
\onlinecite{Beron-Brown-03}%
; in that study, however, attention was confined to ray spreading in $(p,z)$
or, equivalently, $(I,\vartheta ).$ That problem is described by the first
two of Eqs. (\ref{RayTra}) or (\ref{RayTra1}), the upper $2\times 2$ system
in Eqs. (\ref{RayTra2}) and (\ref{S2}), etc. In that study it was shown that
ray spreading is largely controlled by $\omega ^{\prime }.$ The surprising
result of the present study is that travel time spreads are also largely
controlled by $\omega ^{\prime }.$ This is evident from the unconstrained
and constrained travel time spread estimates, (\ref{dT}), (\ref{dTrs}), (\ref%
{CdT}), and (\ref{CdTr}), and the more heuristic arguments associated with
Eqs. (\ref{RayTra2}) and (\ref{S2}).

Inasmuch as the AET experimental observations \cite%
{Worcester-etal-99,Colosi-etal-99} provided much of the motivation for the
present work, it is noteworthy that the results presented here are
consistent with the ray-based analysis of those observations presented in
Ref.
\onlinecite{Beron-etal-03}%
. Two points relating to the results in Ref.
\onlinecite{Beron-etal-03}
deserve further comment. First, it was argued in Ref.
\onlinecite{Beron-etal-03}
that, for moderately steep rays, constrained travel time spreads could be
approximately estimated using $\Delta T_{1},$ the first term on the r.h.s.
of Eq. (\ref{CdTr}). This is moderately surprising in that the AET range was
$3.25$ \textrm{Mm},\textrm{\ }which is large enough that one would expect
that the second term on r.h.s. of (\ref{CdTr}) should be important. The
neglected term, however, is proportional to $\alpha $, which is unusually
small (cf. Ref.
\onlinecite{Beron-etal-03}%
) in that environment for the relevant band of launch angles. Thus the
estimated travel time spread reported in Ref.
\onlinecite{Beron-etal-03}
is close to what one obtains using both terms on the r.h.s. of Eq. (\ref%
{CdTr}). Second, it was noted in Ref.
\onlinecite{Beron-etal-03}%
, without explanation, that in the AET environment constrained travel time
spreads are much larger for near axial rays than for steeper rays. This
behavior is consistent with Eq. (\ref{CdT}) and the observation
\cite{Beron-etal-03}
that in the AET environment $\left| \alpha \right| $ is very large for the
near-axial rays.

The results presented in this paper, coupled with those presented in Refs.
\onlinecite{Beron-Brown-03,Virovlansky-03}%
, represent an important step toward the development of a theory of wave
propagation in random inhomogeneous media (WPRIM). These studies have shown
that both ray amplitude statistics and ray travel time statistics are
largely controlled by the background sound speed profile, via $\omega
^{\prime }.$ It follows that finite frequency wavefield intensity statistics
should also be largely controlled by the background sound speed profile via $%
\omega ^{\prime }.$ This is different from the classical treatment of the
problem of wave propagation in random media (WPRM) which assumes that the
background sound speed structure is homogeneous---for which $\omega ^{\prime
}=0.$ Ideally one would like to develop a uniformly valid theory of WPRIM
that reduces to known WPRM results in the limit $\omega ^{\prime
}\rightarrow 0.$ A more modest goal is to develop an approximate theory of
WPRIM, appropriate for long-range underwater sound propagation, that treats
the case where the wavefield intensity statistics are largely controlled by $%
\omega ^{\prime }.$ To develop such a theory the results presented here and
in Refs.
\onlinecite{Beron-Brown-03,Virovlansky-03}
have to be combined and extended. Necessary extensions are the inclusion of
finite frequency effects (interference and diffraction) and a more accurate
treatment of the link between sound speed perturbations and perturbations to
$I$ and $\vartheta .$ These topics will be explored in future work.

\section*{\textsf{Acknowledgments}}

We thank A. Virovlyansky, J. Colosi, S. Tomsovic, M. Wolfson, G. Zaslavsky,
F. Henyey, and W. Munk for the benefit of discussions on ray dynamics. We
note, in particular, that A. Virovlyansky independently derived Eq. (\ref%
{CdT}); S. Tomsovic independently derived Eq. (\ref{dTrs}) and the first
term on the r.h.s. of Eq. (\ref{CdTr}); W. Munk, J. Colosi, and F. Henyey
independently derived the second term on the r.h.s. of Eq. (\ref{CdTr}),
although not in the action--angle form given here; and F. Henyey and J.
Colosi independently derived Eq. (\ref{dTa}) but not in the action--angle
form given here. Also, the technique of staggering the starting range in a
single realization of an internal wave field to generate effectively
independent realizations was pointed out to us by F. Henyey. This research
was supported by Code 321OA of the Office of Naval Research.

\end{document}